\documentclass{aa}
\usepackage{graphics}
\usepackage{times}
\newcommand {\xbar} {\overline {x}}
\newcommand {\ybar} {\overline {y}}
\newcommand {\bigcr} {\\ \noalign{\medskip}}
\newcommand {\smallcr} {\\*[-5pt]}

\begin{document}
\thesaurus{4(11.11.1; 02.03.1;  05.03.1)}
\title{ Imperfect fractal repellers and irregular families of periodic orbits
                    in a 3-D model potential }

\author{     B. Barbanis, H. Varvoglis and Ch. L. Vozikis}
\institute{Section of Astrophysics, Astronomy and Mechanics, Department
of Physics, Aristotle University of Thessaloniki, 540 06 Thessaloniki,
Greece}
\offprints{H. Varvoglis}
\date{Received 2 July 1998 / Accepted 1 February 1999}

\authorrunning{B. Barbanis et al.}

\titlerunning{Imperfect fractal repellers and irregular families of periodic orbits in a 3-D model potential}

\maketitle

\begin{abstract}
A model, plane symmetric, 3-D potential,
which preserves some features of galactic problems,
is used in order to examine the phase space
structure through the study
of the properties of orbits crossing perpendicularly the plane of
symmetry. It is found 
that the lines formed by periodic orbits, belonging to Farey sequences,
are not smooth neither continuous.
Instead they are deformed and broken in regions characterised by
high Lyapunov Characteristic Numbers (LCN's).
It is suggested that these lines are an incomplete
form of a fractal repeller, as discussed by Gaspard and Baras (1995),
and are thus closely
associated to the ``quasi-barriers'' discussed by Varvoglis et al. (1997).
There are numerical indications that the contour lines of constant
LCN's possess fractal properties. Finally it is shown numerically
that some of the periodic orbits -members of the lines- belong to true
irregular families.
It is argued that the fractal properties of the phase space should affect 
the transport of trajectories in phase or action space and,
therefore, play a certain role in the chaotic motion of stars in
more realistic galactic potentials.

\end{abstract}


\section{Introduction - Motivation}
     One  of the most interesting open questions in non-linear
dynamics is the nature and evolution of transport in the
chaotic phase space regions of perturbed integrable Hamiltonian systems. Most
of the work in this field has been done for 2-D systems or
for the standard map, following two different
approaches. In the first approach, which may be designated as
{\it microscopic}, one studies the homoclinic and heteroclinic tangle
of stable and unstable
manifolds inside the chaotic region. This gives a ``complete''
qualitative picture of transport in phase space, since a chaotic
trajectory follows the unstable manifolds of the unstable
periodic orbits, jumping from one to another near the
heteroclinic points. On the other hand, invariant tori surrounding
stable periodic orbits act as ``barriers'' in the transport
process, modifying at the same time the topological structure of
phase space. However, it is not presently well understood how the
results from this kind of study can be used in a quantitative
description of the evolution of an ensemble of trajectories.

     In the second approach, which may be designated as {\it
macroscopic}, one assumes {\it a priori} that transport is a pure
random phenomenon (a Markovian process). Then the transport of an
ensemble of trajectories may be considered as ``classical
diffusion'' in action space.
Unfortunately, it is well known that transport in Hamiltonian
systems cannot be considered as a pure Markovian process,
due to the problem of
{\it stickiness}. Moreover the process may follow L\'evy rather
than classical statistics (Shlesinger et
al., 1993). In this case the macroscopic approach still works,
provided that the process is considered as {\it fractal} diffusion
in {\it normal} space (Zaslavsky, 1994). In this way it becomes evident that
L\'evy-like statistics are closely related to the self-similarity of
the phase space. This formalism results in a differential equation
with fractional partial derivatives, whose solution is formidably
difficult, even in the simple case of 2-D systems.

     There are rather few attempts to study transport in
Hamiltonian systems with more than two degrees of freedom, mainly
because the above mentioned approaches for 2-D systems cannot be
directly generalized. It is therefore important to note
that there exists yet a third approach for the description of
transport, which may be implemented in a straightforward way in
systems with more than two degrees of freedom, and this is {\it
normal diffusion} in {\it fractal space} (e.g. see West and
Deering, 1994, and references therein). The fractality of phase space
has been already demonstrated for 2-D systems and has been
associated with the self similarity of the hierarchical
structure of island families on a surface of section (e.g. see
Zaslavsky, 1994, BenKadda et al., 1997). 

In the case of more than
two degrees of freedom the situation is more complicated, since
the topological structure of phase space in any region depends on
the number of local integrals of motion. Recently
Varvoglis et al. (1997) have found indications
of phase space fractality in the same model 3-D Hamiltonian
system studied in the present work, but they were unable, due to the
particular method of study they had selected,
to identify actual fractal or multifractal sets in phase space.
They have conjectured that the fractality is due to the
presence of {\it quasi-barriers} (Varvoglis and Anastasiadis,
1996), i.e. geometrical objects of lower than full dimensionality
(depending on the  number of local integrals of motion), such as
periodic orbits, invariant tori around them and cantori (Aubry,
1978; Percival, 1979). 

From the above discussion it is obvious that the importance of the
fractality of phase space in the
diffusion process lies in the nature and in the rate of the diffusion:
In a simply connected space,
the diffusion is classical and follows Fick's law (transport
proportional to $\sqrt (t)$). In a fractal
space the diffusion is non-classical (presumably a L\'evy process)
and, in the case of a
Hamiltonian system, usually it has a lower than the Fickian rate.

On the other hand, the distribution of
periodic orbits perpendicular to the $x-y$ plane
in the present 3-D model potential
has been studied by Barbanis and Contopoulos (1995) (henceforth referred
to as BC) and Barbanis (1996). It was found that it is not
random: the perpendicular crossings (p.c.'s)
of orbits with multiplicities forming Farey sequences with the $x-y$ plane
are arranged along lines,
named in the present paper basic-lines (BL) and Farey tree lines (FTL).
These lines are not continuous but
they possess gaps; two of them have already been mentioned in BC.
Since the presence of Farey sequences
implies a form of self-similarity, it is possible that the BL's and FTL's
may be related to the above-mentioned quasi-barriers and that the gaps
may be related to any fractal properties of phase space.

In the present paper we use the model potential studied by
BC in order to test the conjecture put forward
by Varvoglis et al. (1997) about the origin of the fractality of phase space.
Furthermore, through this, we attempt to assess
the relation between
the topological structure of the phase space and transport, on the one hand,
and the system of BL's and FTL's, on the other.
Although this potential can model a true galaxy only locally,
its study is expected, nevertheless, to contribute to the understanding
of galactic evolution problems, based on the fact that evolution is intimately
connected to transport in phase and configuration space.
Transport in a chaotic region of a perturbed 
integrable dynamical system, in turn, may be viewed as
(stochastic) diffusion, which depends on the value of the
Lyapunov Characteristic Number as well as on the topological structure of the
phase space. We believe that similar
behaviour would characterise any 3-D perturbed integrable dynamical system,
as are most of the realistic model galactic potentials.

The
purpose of this paper is threefold:

(a) to study in finer detail the
distribution of the periodic orbits along the BL's and FTL's.

(b) to examine whether these lines are connected, in any way,
to transport in phase space and are, thus, related to the
``quasi-barriers'' and

  to confirm the existence of true
irregular families of periodic orbits in a 3-D Hamiltonian system.

     This paper is organized as follows. In the next section we
present, for reasons of completeness, some remarks on
previous results. In Section 3 we show that the LCN's may ``map''
the fractality of phase space. In Section 4 we present our
results on the interconnection and discontinuities of BL's and FTL's.
In Section 5 we show numerically the existence of irregular families
of periodic orbits. 
Finally in Section 6 we summarise and discuss our results.

\section{Remarks on previous results}

\subsection{Basic lines}
\begin{figure}
  \resizebox{\hsize}{!}{\includegraphics*{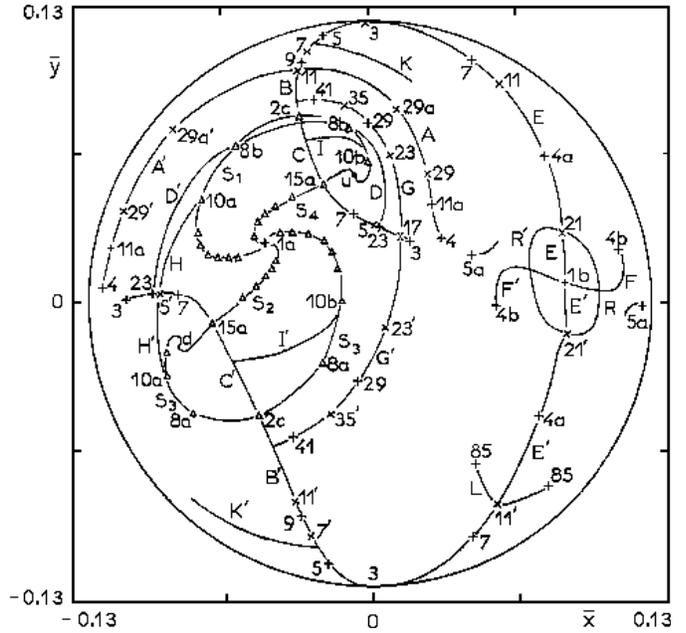}}
  \caption{The majority of the perpendicular crossings
  of the periodic orbits of the Hamiltonian (1) lie on particular lines on
  the $\xbar-\ybar$ plane inside  the boundary circle
  $\xbar^2+\ybar^2=2h$. Most of the lines are concentrated around
  the basic orbits
  1a and 1b. There is also a noticeable  spiral structure with focal
  point the orbit 1a. Four spirals  illustrate the area of this 
  structure. Note that in this figure (reproduced from BC, Fig. 5) a different
  notation is used than
  in the rest of the present paper. Each
  symbol, $(\times)$ or $(+)$, shows a periodic orbit with one or two
  crossing points, respectively.} 
  \label{figM1}
\end{figure}

     The systematic exploration of the phase space structure of
3-D Hamiltonian systems was initiated by Magnenat (1982) and later
by Contopoulos \& Barbanis (1989), BC and Barbanis (1996),
through the study of periodic orbits of unit mass test
particles in a 3-D model potential, corresponding to the Hamiltonian
\begin{eqnarray}
    H = & {1 \over 2} \left( p_x^2 + p_y^2 + p_z^2 \right) +
    {1 \over 2} \left( A x^2 + B y^2 + C z^2 \right) \\ 
    & {-\epsilon x z^2 - \eta y z^2 = h} \nonumber
\end{eqnarray}
with a particular set of parameters (A = 0.9, B = 0.4, C =
0.225, $\epsilon = 0.560$, $\eta = 0.20$ and $h = 0.00765$). The
specific values for the parameters A, B, C were
selected by Magnenat (1982) because the 2-D system $x-z$ for
$\sqrt{A/C}$ = 2 is topologically
equivalent to the Inner Lindblad Resonance and the 2-D system $y-z$ for
$\sqrt{B/C}$ = 4/3
is a well studied 2-D dynamical system with extended chaotic regions.
For this set of parameters it was found in
BC that the distribution of the p.c.'s
of the periodic orbits is not accidental: the crossings are arranged along
particular lines on the $\xbar-\ybar$ plane (where $\xbar =
A^{1/2}x$, $\ybar = B^{1/2}y$), as it is evident in Fig.~\ref{figM1}
(taken from Fig.~5 of BC). Each orbit has either one p.c., marked
by a (x), or two p.c.'s, marked by (+). In this paper, for clarity
reasons, we use a point ($\cdot$) instead of (x). Each p.c. is
designated by a number (or a number and a letter), representing
the multiplicity of the orbit; different orbits with the same
multiplicity are differentiated by a  prime. The lines are formed
by the p.c.'s of the orbits, whose multiplicities form arithmetic
progressions with increment the multiplicity of the limit orbit.
They are termed in this paper {\it basic} if they connect orbits
of the unperturbed system (e.g. 1a, 1b in Fig. 1) or orbits with low
multiplicity, $m$ (e.g. $m \leq 5$).  For example the basic
sequences
\begin{itemize}
\item[] \hspace{0.5cm} 3(+), 5($\cdot$), 7(+), 9($\cdot$), ...2c(+) 
\item[] and 3(+), 5$'$($\cdot$), 7(+), 9$'$($\cdot$), ...2c(+),
\end{itemize}
which form the lines C and C$'$, start with
an orbit of multiplicity 3 and increment 2, which is the
multiplicity of the limit orbit 2c.

\subsection{Farey tree lines}
     In BC it was recognized that the distribution of p.c.'s on the
$\xbar-\ybar$ plane presents some self-similar features. To begin with,  to
each basic sequence of orbits corresponds a number of new sequences of higher
order. Each member of a new sequence is a periodic orbit having
as multiplicity the sum of the multiplicities of two consecutive
orbits of the generating sequence. For example between the first
two orbits of the basic sequence C, i.e. 3(+) and 5($\cdot$), there is
the orbit 8(+). Two second order sequences are formed between 3
and 8 as well as between 8 and 5, i.e.
\begin{eqnarray}
   & 8(+), 11(\cdot), 14(+), 17(\cdot), 20(+), 23(\cdot), ...3(\cdot)
\nonumber\\ 
\mathrm{and} & \\
 & 8(+), 13(+), 18(+), 23(+), 28(+), ...5(\cdot)\nonumber 
\end{eqnarray}
Both new sequences have as first member the orbit 8, but they
have different limit orbits and increments (i.e. 3 for the first
and 5 for the second, respectively). In the same way we can form
sequences of even higher order.
We call the orbits of these higher order
sequences {\it Farey tree orbits} and the lines on which
they lie {\it Farey tree lines}, because such sequences are
similar to the Farey sequences discussed by Niven and Zuckerman (1960).
For a good review on Farey trees see Efthymiopoulos et al. (1997). 

The periodic orbits are used, in the present paper, as an additional tool
for probing the
topological structure of the phase space. In particular, periodic
 orbits of high multiplicities are
needed in order to search for self-similar structures in the BL's 
and FTL's for a wide interval of
multiplicities (see Avnir et al., 1998). The importance of the 
non-basic (Farey) lines lies exactly
on the above line of reasoning, as well as in the fact that they
might be necessary in order to
calculate correctly an appropriate diffusion coefficient. It should 
be noted that we have restricted
our study to the periodic orbits crossing perpendicular the x-y plane
for simplicity reasons.

\begin{figure}
\resizebox{\hsize}{!}{\includegraphics*{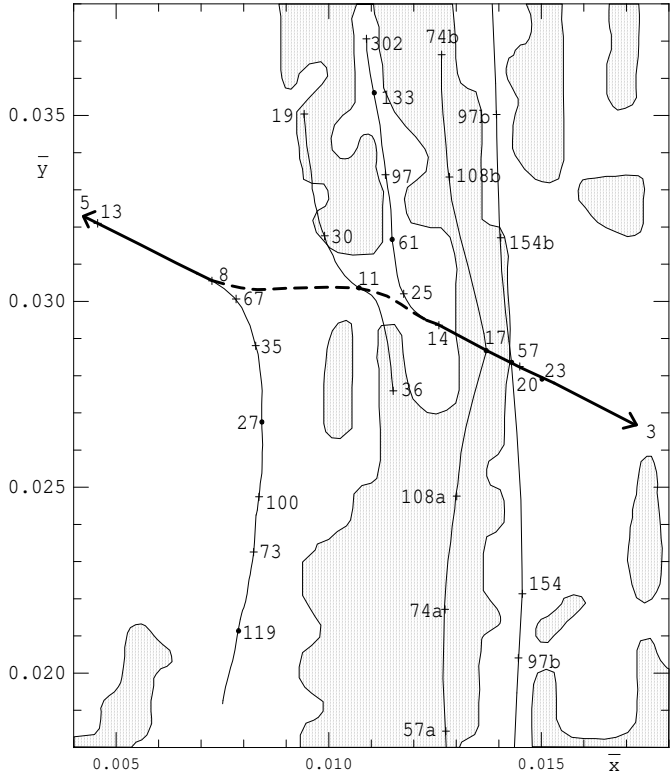}}
\caption{A segment of the BL between the p.c.'s of the periodic
orbits 3 and 5 (indicated by a bold line) and the network of various FTL's
(corresponding to the upper central part of Fig.~\ref{figM1}, line C).
The nonexistent part of C between 8 and 14
is indicated by a dashed line. The contour lines
of constant LCN's  $\lambda = 10^{-4}$ are superimposed and the
regions of $\lambda  > 10^{-4}$ are gray shaded}
\label{figM2}
\end{figure}

\subsection{Irregular orbits}
All periodic orbits of the same multiplicity that can be found through the
continuous variation of a parameter in the equations of motion belong
to the same 
family. The graph giving one of the initial conditions of the periodic
orbits as a function of the parameter is the {\it characteristic}
of this family. Families, which do not exist for values of the
perturbing parameter below a critical value, appear only as pairs, are well
known in 2-D dynamical systems and are called {\it irregular
families}  (Contopoulos 1970; Barbanis 1986).
However, their existence has not been proven for 3-D systems.
Pairs of families that have been found in such systems without
a direct connection to a basic orbit, they were all of small
multiplicities and  
it turns out that they have an indirect connection to a basic orbit
through other families bifurcating from it 
(Barbanis, 1996).
It is desirable to know whether
families of periodic orbits in 3-D systems (presumably of high multiplicities)
do exist without {\it any} connection to a basic orbit.
Their presence might play a certain role in the fractality of phase
space and, therefore, 
in trajectory transport.

\subsection{Lyapunov Numbers}
Varvoglis et al. (1997), while
studying numerically transport phenomena in the trajectories of
the Hamiltonian (1), have found, in an indirect way, that the phase
space shows a 
fractal structure. In particular they presented numerical
evidence that the function $dV/dt$ has multifractal properties,
where $V(t)$ denotes essentially the coarse-grained volume of phase
space visited by a trajectory up to
time $t$. They found, also, that the evolution of the function $dV/dt$
is closely 
related to the evolution of the function 
\begin{equation}
\chi (t) = { 1\over t} \ln {d(t)\over d(0)}
\end{equation}
whose limit, as $t \rightarrow \infty$, is defined as the LCN. The
function $\chi (t)$ shows plateaus in the time intervals where $dV/dt$
is close to 
zero and steeply increasing transient
segments in the time intervals where $dV/dt$ takes large positive
values. This fact shows that a trajectory is confined successively
in regions of phase space where the LCN converges to a limit and, therefore,
the fractal properties of $dV/dt$ should arise from the
(apparently) self-similar distribution of periodic orbits and
other quasi-barriers in phase space (Zaslavsky, 1994).

It should be emphasised that, in the present paper, LCN's are not
calculated in order to estimate
``rates of diffusion'', since LCN's alone
cannot describe completely this phenomenon in the case where there exist
invariant tori of considerable measure. We do,
however, calculate LCN's in 
order to ``probe'' the phase space structure, i.e. to delineate, in an
independent from other methods 
way, the ``topology'' of the phase space. This new method is based on
the property that 
trajectories starting in a certain region, bounded by
``quasi-barriers'', are characterised by a certain 
``Local Lyapunov Number''. By drawing the
contour lines of LCN's, one 
draws, essentially, the geometric boundary of the region bounded by
the ``quasi-barriers''.

\section{LCN's and fractal properties}

The ``classical'' calculation of LCN's
requires a continuous integration of the corresponding
trajectories, until the function $\chi (t)$ has
reached a plateau. It is, therefore, obvious that this method cannot
be used in a ``mass production''
procedure. For this reason we decided to use a different approach, by
calculating the value of the function $\chi (t)$
at various integration time intervals, $\Delta t$, up to $\Delta t = 3\,10^5$.
Comparing the results for different $\Delta t's$ we have found that they
are qualitatively the same (i.e. they form level lines with increasing
complexity as one goes to finer details),
provided that $\Delta t > 10^5$.
So, even if $\chi (t)$ does not reach a plateau, its value at the end of the
corresponding time interval $\Delta t > 10^5$
gives a correct estimate, at least to order of magnitude, of the
``degree of stochasticity'' of the trajectory in the phase space region
restricted by the quasi-barriers. We designate this value
as the {\it Fixed Time LCN}. In what follows we use, for
convenience, the notation LCN. Through this kind of LCN's we attempt to
probe the topological structure on the $\xbar-\ybar$ plane as
follows:

We draw a dense mesh on the $\xbar-\ybar$ plane (for $0 < \xbar <
0.020$ and $0.010 < \ybar < 0.045$ at intervals $\Delta \xbar$ =
$\Delta \ybar = 0.0005$) and we use the nodes of this mesh as
initial conditions for trajectories starting perpendicular to this plane.
We calculate  the LCN of each trajectory at the end of various
time intervals and we draw the contour lines of constant LCN
on the $\xbar - \ybar$ plane at the LCN value $\lambda = 10^{-4}$. In
Fig.~\ref{figM2} 
we show our results for the LCN's at
the end of a time interval $\Delta t = 3\,10^5$.
Fig.~\ref{figM2}  contains only a small part of the studied region. The
structure of the 
contour lines in the whole region as well as their variation with $\Delta t$
is still under investigation and
will be the topic of another publication.

The emergent
picture is very interesting. One can immediately see that the
studied area may be divided into two regions: Most of it is
characterized by small values of LCN's (white regions);
inside the white regions one
can find small elongated regions characterized by high values
of LCN's (gray regions).
It should be noted that the position and elongation of these
regions is closely related to the position and the direction of BL's and
FTL's (see next Section). This fact may be interpreted in the
following way. Since a considerable fraction of the periodic orbits
along these lines are unstable, the exponential
divergence of 
trajectories in the area around them is governed mainly by the
unstable manifolds of the orbits-members of the lines. The situation
is reminiscent of a {\it fractal repeller} as defined in Gaspard and
Baras (1995), i.e. a set of countably 
infinite unstable periodic orbits, which has zero Lebesgue measure
and a finite Hausdorff
dimension. In our case the orbits-members of BL's and FTL's may be
considered as forming a sort of an incomplete
(since not all periodic orbits are unstable) {\it fractal repeller}.
In other words
the BL's and FTL's are closely associated with the ``quasi-barriers'',
which are geometrical objects that separate ``loosely'' different
phase space regions with 
different Local LCN, as it has been discussed 
by Varvoglis et al. (1997) and Tsiganis et al. (1998).

\begin{figure}
\resizebox{\hsize}{!}{\includegraphics*{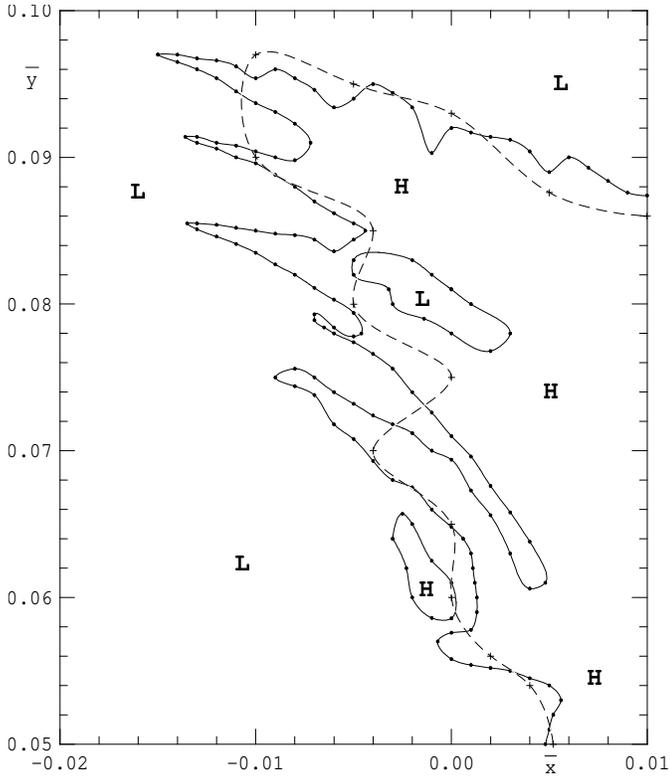}}
\caption{The contour lines of constant LCN at the value $\lambda = 10^{-4}$
for a coarse
mesh ($\Delta \xbar = \Delta \ybar = 0.005$, dashed line) and a fine mesh
($\Delta \xbar = \Delta \ybar = 0.001$, solid line). Regions of
 $\lambda > 10^{-4}$
and $\lambda < 10^{-4}$ are marked with a H and an L, respectively}
\label{figM3}
\end{figure}

Since (I) there is evidence that BL's and FTL's have 
fractal properties and (ii) there is a relation
between the BL's and FTL's, on one hand, and the
LCN's on the other, it should be interesting to examine
whether the contour lines have fractal properties as well. To do so we
calculate the LCN's in the region $-0.02< \xbar < 0.01, 0.05 < \ybar < 0.10$
using a coarse ($\Delta \xbar = \Delta \ybar = 0.005$)
and a fine mesh ($\Delta \xbar = \Delta \ybar = 0.001$). This
region was selected because it was studied, although with a much coarser mesh
( $\Delta \xbar = \Delta \ybar = 0.01$)
by Contopoulos and Barbanis (1989, Fig. 9).
We then plot
both contour lines at $\lambda = 10^{-4}$, on the same graph, as
shown in Fig.~\ref{figM3}.
It is easy to see that, while the dashed line, corresponding to the
coarser mesh, seems rather smooth, the continuous one, corresponding to the
finer mesh, shows ``tongues'' that oscillate
about the dashed curve, a picture reminiscent of the classical
examples of fractal curves. Of course this cannot be taken as a {\it
proof}, even numerical, that contour lines are fractal curves, since
the available data are not exhaustive (see also Avnir et al.,
1998). However, we believe that Fig.~\ref{figM3}, if considered
together with the 
fact that LCN's reflect the distribution of the unstable manifolds of
unstable periodic orbits, is a strong indication that the contour
lines have, indeed, fractal properties. 

\section{BL and FTL are not simple}

As it is evident from Fig.~\ref{figM1} and the relevant discussion in BC
and Barbanis (1996), the BL's formed by the
lowest order sequences look continuous and smooth.
Because of this it had been assumed that the
lines appearing in Fig.~\ref{figM1} contain all the higher order
sequences described in Section 2 as well, so that any fractal properties,
arising from the Farey tree character of the p.c. sequences, are
restricted {\it along} the lines and not {\it across} them. However,
the existence of two small
gaps in a BL, noticed by BC, gave the first evidence that, somehow,
this picture is 
not correct. In order to examine, therefore, in depth the
situation, we proceed in the calculation of
periodic orbits belonging to several higher order sequences than those
appearing in
Fig.~\ref{figM1}.

We focus our interest on a small region, between the orbits 3(+) and
5($\cdot$) of the 
line C (BC, Fig. 8), since it is in this area that the first
noticeable gap was observed.
Between these two orbits lies
the orbit 8(+). Therefore two second order sequences are formed with orbit
8 as first member and increments 3 and 5 respectively, i.e.
the sequences (2). In Fig.~\ref{figM2} we can see that the
first sequence 
presents a large gap between orbits 8 and 14, where higher
order FTL's, branching away from the BL, may be observed.
Furthermore, as we discuss below, orbit
11 (which lies between orbits 8 and 14, instead of being part of the BL)
belongs to a higher order FTL. In contrast the second
sequence does not show any noticeable gaps.
The reason for this difference becomes obvious, if one
considers the BL's and FTL's in association with the degree of stochasticity
of the phase space where these orbits lie, estimated through
the calculation of LCN's.
In Fig.~\ref{figM2} one can immediately notice
that the first sequence crosses a region of high LCN's,
while the second is confined in a region of low LCN's.

Let us now examine what happens to the higher order sequences
between 11 and 8, on one hand, and 11 and 14 on the other, i.e.
11($\cdot$), 19(+), 27($\cdot$), 35(+), ..., 8(+) and
14(+), 25(+), 36(+), ..., 11($\cdot$).
As it is evident from Fig.~\ref{figM2},
the FTL corresponding to the first sequence is torn and split into
two branches, one beginning from
orbit 11 going upwards and the other beginning from orbit 8
going downwards.
Similar is the situation with the second sequence. The FTL is torn and split
into two branches, one going upwards from orbit 14 and the other
going downwards from orbit 11. The branch going downwards from orbit 11
is connected to the FTL
of the first sequence going upwards from orbit 11, resulting in a
composite S-shaped line (the three leftmost continuous
thin lines in Fig.~\ref{figM2}). 
In this way we see that 
the gap between orbits 8 and 14 is, in a way, ``filled''
with FTL's corresponding to higher order sequences. This is an example
of a case where 
higher order FTL's do not lie on the ``parent'' BL. 

It is interesting to note that the branch of the second FTL
going upwards from orbit 11 is developed inside a
narrow strip of ordered motion.

In Fig.~\ref{figM2} we notice also two ``vertical'' higher order FTL's
that emanate from the orbits 57($\cdot$) and 17($\cdot$), corresponding
to higher order sequences with increment 40. The first of them,
the one emanating from
orbit 57($\cdot$), is vertical to the local gradient of the contour lines
and lies in a region of low LCN's. The second one, the one emanating
from orbit 17($\cdot$), is vertical to the local gradient of the contour lines
as well but it seems to span a {\it high}
LCN's region. The most probable explanation is that the line
lies, in fact, in a narrow strip of low LCN's, with a width less
than the size of the mesh used. This is something that should
be expected for a geometrical object with fractal properties
and has been indeed encountered in some other
regions of the studied area.

From the above discussion emerges an intuitively appealing picture:
BL's and FTL's are, in a sense, lines representing some ordered features of the
dynamical system, which are deformed in the neighbourhood of the
{\it chaotic seas}
and tend to run perpendicular to the local gradient of the
contour lines of constant
LCN's. Therefore we understand that the structure of
the BL's and FTL's on the  $\xbar - \ybar$ plane is considerably more
complicated than it was assumed in BC, a fact which, as we showed in Section 3,
seems to play an important role in the nature of transport in the phase
space of Hamiltonian (1).

\section {Existence of irregular families}

In two previous papers (BC; Barbanis 1996) the authors investigated
the bifurcation and the evolution of known periodic orbits belonging to some
BL's. It was found that a small number of these families
form pairs that do not have any direct or indirect connection with the 
periodic orbits of the unperturbed system. However, for a given multiplicity,
there are many other bifurcating families, which may play the role of
connecting bridges between those pairs.

Following these considerations, we studied the evolution of 
the bifurcations of some {\it specific} multiplicities from the
basic orbits 1a and 1b. We have selected the multiplicities 11
and 29 for three reasons: (I) 11 and 29 are prime numbers, so that
there are no families of these multiplicities resulting from lower
multiplicity families, except from those with multiplicity one.
(ii) The number of bifurcating families is neither small nor
great. (iii) There are a lot of orbits of these multiplicities on
the various BL's and FTL's for $\epsilon = 0.560$ and $\eta = 0.20$.

The detailed discussion in Appendix A and the examples in Appendix B lead to
the conclusion that pairs of families 
without any direct or indirect connection to a basic orbit of the
unperturbed system (e.g. the six pairs
appearing in Fig.~\ref{figA1}) exist in a 3-D system. This happens
mainly for high multiplicities, exactly
as one would expect considering the analogous case of irregular families
in 2-D systems. It should be noted also that, besides the above mentioned
{\it classical} case of irregular families, in this work we have found a new
kind of irregular families, which we call a {\it double pair}, i.e
a pair consisting of two single pairs (see Fig.~\ref{figB6} and its discussion
in Appendix B).

\section{Summary and conclusions}
Barbanis and Contopoulos (1995), in studying the
distribution of the p.c.'s of the periodic orbits in a well known 3-D model
potential with the 
$\xbar-\ybar$ plane, have discovered a noteworthy order. Except for a
couple of gaps, the p.c.'s
of periodic orbits, whose multiplicities form Farey-tree sequences,
are arranged 
along continuous and smooth-looking lines. The p.c.'s along these
lines have some fractal 
properties, arising from the fact that they belong to Farey-tree sequences
of various orders. In the present work we tried to study the properties
of these lines in association with the values of the LCN's of the area
span by them.

We have found numerical evidence that the contour lines of 
LCN's
show fractal properties. We have also found numerical evidence that the
splitting of initially continuous and smooth-looking 
BL's and FTL's, as well as the ensuing formation of gaps, is not an exception,
as was implied in BC. As a rule, it is observed in regions
of the $\xbar - \ybar$ plane characterised by high values of the LCN's and
appears not only in BL's, but in higher order
FTL's as well.

Since our results are only numerical and involve only a small
number of
periodic orbits, the above results cannot be considered as firm proof
that the distribution of BL's and FTL's as well as the contour lines
of LCN's have, beyond any doubt, fractal properties.
However we feel that there is enough evidence that the above two
geometrical objects 
(BL's and FTL's on the one hand and contour lines on the other)
are closely related and that, furthermore, they show strong indications
of fractal behaviour. Therefore we think that the study of some
selected regions of the $\xbar - \ybar$ should be done in more detail,
by calculating LCN's in even finer mesh
and by comparing the contour lines to higher order FTL's,
in order to establish to a higher level of confidence the fractal
nature of both geometrical objects.

As far as irregular periodic orbits are concerned, we have found
that the majority of the periodic orbits forming BL's and FTL's either
bifurcate directly from 1a or 1b or they are connected indirectly with
them. However, 
by considering families of periodic orbits with large multiplicities,
we find that some  
of them do not have any connection with 1a or 1b but they form a pair with
another family. In the present work we have encountered also the more complex
situation of a {\it double pair}, i.e. a pair whose members are pairs too.

\begin{acknowledgements}
The authors would like to thank Prof. L. Martinet for a critical
reading of the first version, which improved the final form of the
paper. H.V. would also like to thank Prof. J.H. Seiradakis and
Mr. K. Tsiganis for 
several helpful discussions.
\end{acknowledgements}

\appendix

\section{Search for irregular families; bifurcations from the orbit 1a}

\begin{figure}
  \resizebox{\hsize}{!}{\includegraphics*{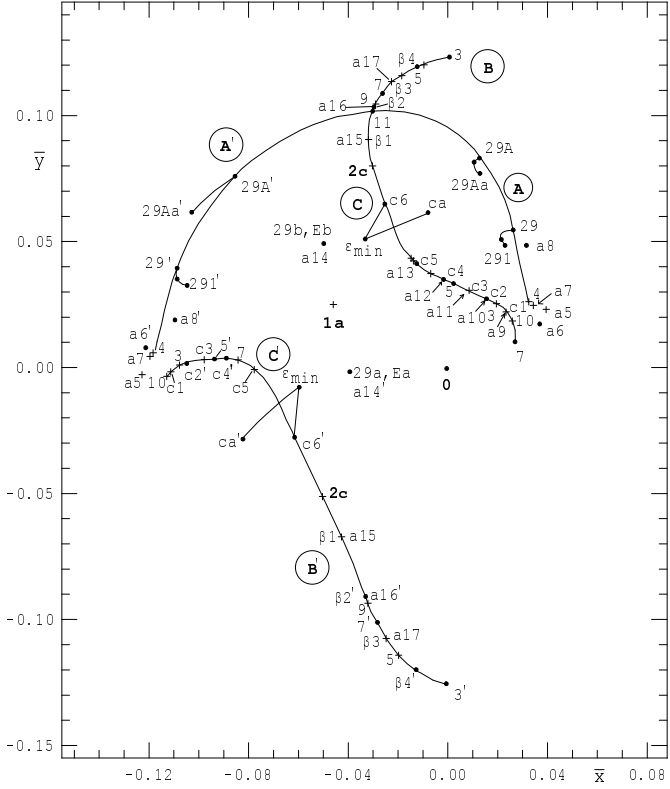}}
  \caption[]{ The lines A and A$'$,
B and B$'$, C and C$'$ are around orbit 1a. 
 On these lines the p.c.'s of the periodic orbits with multiplicity
29 are depicted, when $\epsilon=0.560$,
$\eta=0.20$. The orbits of the corresponding families of the 
bifurcations from orbit 1a of Table A1 are also given, omitting
  from each name the prefix 29.} 
  \label{figA1}
\end{figure}

\begin{figure}
  \resizebox{\hsize}{!}{\includegraphics*{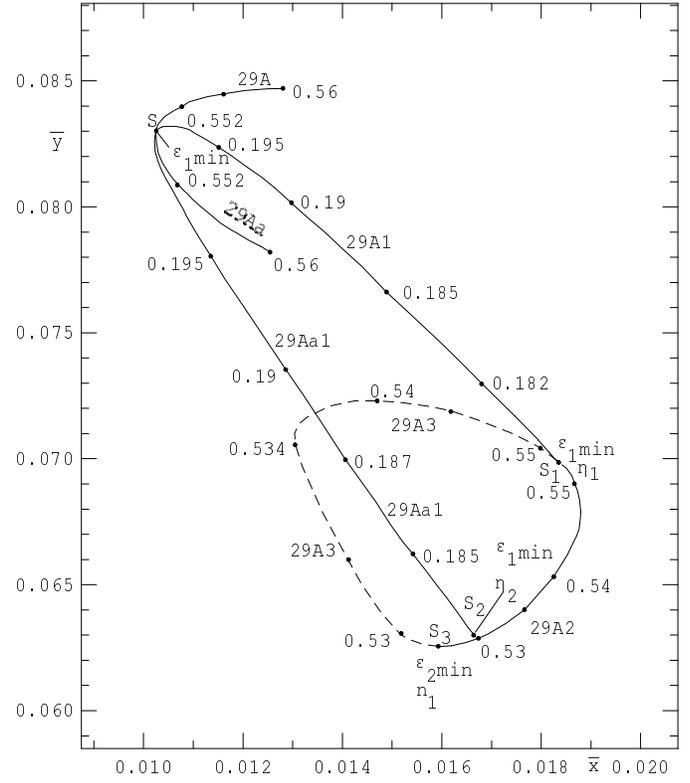}} 
  \caption[]{The family 29A forms a pair with the family 29Aa.
From the common point S ($\epsilon_\mathrm{_1 min},\eta$)
emanate the family 29A1 and 29Aa1 by changing
$\eta$. The family 29A1 stops at the point S1 
($\epsilon_\mathrm{_1 min},\eta_1$), 
while 29Aa1 at the point S2 ($\epsilon_\mathrm{_1 min},\eta_2$).
No continuation of 29Aa1 has been found. On the other hand at the point S1
two other families, i.e. 29A2 and 29A3 emanate with $\eta = \eta_1$
and changing $\epsilon$ between $\epsilon_\mathrm{_1 min}\leq \epsilon
\leq \epsilon_\mathrm{_2 min}$. The two families are connected at the point
S3 ($\epsilon_\mathrm{_2 min},\eta_1$). No connection of  S3 with 1a
has been found.}
  \label{figA2}
\end{figure}

     Tables A1 and A2 give the bifurcating families
from the basic family 1a with multiplicities 29 and 11 respectively,
when $0 \leq \epsilon \leq  0.560$ and $\eta = 0.20$. The corresponding
bifurcations from 1b are given in Appendix B. The bifurcations
from 1a and 1b take place when the index
\begin{equation} 
 a = -2 \cos \left( 2 \pi R \right) 
\end{equation}
(where $R$ is the rotation number of the bifurcating orbit) is
equal to one of the two indices of the stability, $s$, of
the orbits 1a or 1b (Contopoulos \& Barbanis 1994). In this way we
calculate $\epsilon_\mathrm{bif}$, i.e. the value of the parameter $\epsilon$ 
(for $\eta = 0.20$) for which one of the indices of the stability of
the orbit 1a or 1b, calculated in quadruple precision, is equal
to the index $a$.

Each Table contains  the
rotation number, $R$, the index
$a$, the values of $\epsilon_\mathrm{bif}$,
the name of each bifurcating family and the way
that this family evolves.
The name of each bifurcating family from 1a or 1b consists
of a letter (a or b) and two numbers. The first number gives 
the multiplicity, $m$ (11 or 29), and the second
the nominator of $R$.
For example, 11a7
designates the family with multiplicity 11 which bifurcates from
1a and has $R$ = 7/11.
Orbits of families whose nominator of
$R$ is odd have two p.c.'s with the ($\xbar-\ybar$) plane, while those with
even nominators have only one. In the case of one p.c. there are
two different bifurcating families. We distinguish one of them
from the other with an accent to its family name, i.e. 11a6$'$.

In Fig.~\ref{figA1} we have drawn the BL's A and A$'$, B and B$'$, C and C$'$.
The origin and the end of these lines are the orbits 4(+) and
11($\cdot$) for A and A$'$, 3($\cdot$) and 2c(+) for B, 3$'$($\cdot$)
and 2c for B$'$, 3(+) and 2c(+) for C and C$'$.
For clarity reasons we omit the prefix 29 from the name of the
families in Fig.~\ref{figA1} and Table A1.

\begin{table}
\caption[]{Bifurcations from 1a with m=29}
\begin{flushleft}
\begin{tabular}{lllll}
\hline\noalign{\smallskip}
R & $a$ & $\epsilon_\mathrm{bif}$ & Fam. & Comments \\
\noalign{\smallskip}
\hline\noalign{\smallskip}
1/29& -1.95324111 & 0.046714 & a1 & highly unst.
\bigcr
&&& a2 \smallcr
2/29 & -1.81515084 & 0.104470 &  &   $\gg$ \smallcr
& & & a2$'$
\bigcr
3/29 & -1.59218613 & 0.157357 & a3  & $\gg$
\bigcr
&&&a4 \smallcr
4/29 & -1.29477257 & 0.207020 &  & $\gg$ \smallcr
&&&a4$'$
\bigcr
5/29 & -0.93681688 & 0.253711 & a5  & see points on \smallcr
&&&&Fig.~\ref{figA1}
\bigcr
&&& a6\smallcr
6/29 & -0.53505668 & 0.297548 &  & $\gg$\smallcr
&&& a6$'$
\bigcr
7/29 & -0.10827782 & 0.338633 & a7  & $\gg$
\bigcr
& & & a8\smallcr
8/29 & 0.32356399 & 0.377069 & & $\gg$\smallcr
& & & a8$'$
\bigcr
9/29 & 0.740277631 & 0.412949 & a9  & a9=c1
\bigcr
& & & a10  & a10 = c2\smallcr
10/29 & 1.12237413 & 0.446314 &  \smallcr
& & & a10$'$ &  a10$'$ =  c2$'$
\bigcr
11/29 & 1.45199098 & 0.477053 & a11 & a11=c3
\bigcr
& & & a12 &  a12=c4\smallcr
12/29 & 1.71371435 & 0.504489 &  \smallcr
& & & a12$'$  & a12$'$=c4$'$
\bigcr
13/29 & 1.89530634 & 0.525074 & a13 & a13=c5
\bigcr
& & & a14  &a14=29b \smallcr
14/29 & 1.98827591 & 0.5184375&  \smallcr
& & & a14$'$  &a14$'$=29a 
\bigcr
15/29 &  1.98827591 & 0.467095 & a15  & a15=$\beta$1
\bigcr
& & & a16  & a16=$\beta$2\smallcr
16/29 & 1.89530634 & 0.395884 & \smallcr
& & & a16$'$  & a16$'$=$\beta$2$'$
\bigcr
17/29 & 1.71371435 & 0.310435 & a17 & a17=$\beta$3
\bigcr
& & & a18 & a18$\rightarrow\beta$4\smallcr
18/29 & 1.45199098 & 0.208203 & \smallcr
& & & a18$'$ & a18$'$$\rightarrow\beta$4$'$
\bigcr
19/29 & 1.12237413 & 0.085467 & a19 & a19$\rightarrow$a191\\
\noalign{\smallskip}
\hline
\end{tabular}
\end{flushleft}
\end{table}

The known orbits of multiplicity 29 on the above  lines are:
\begin{itemize}
\item On the lines A and A$'$ : 29 and 29$'$, 29A and 29A$'$
\item On the lines B and B$'$ : $\beta$1, $\beta$2 and $\beta$2$'$,
$\beta$3, $\beta$4 and $\beta$4$'$
\item On the lines C and C$'$ : c1,c2 and c2$'$, c3, c4 and c4$'$, c5,
c6 and c6$'$ 
\item On other FTL lines (not shown here) lie the termination points 
of the bifurcating families from 1a, with $R$=5/29 to 8/29, namely a5, 
a6 and a6$'$, a7, a8 and a8$'$ (see Table A1)
\end{itemize}

The families of Table A1, which correspond to the first four rotation numbers,
become 
highly unstable as $\epsilon$ increases. This is the reason why
the computations were not continued until $\epsilon = 0.560$.

The corresponding orbits of the families with $R$=9/29 to 13/29,
when $\epsilon = 0.560$, are the points c1, c2 and c2$'$, c3, c4
and c4$'$, c5 of C and C$'$ respectively. Similarly, the orbits of the
families with $R$=15/29 to 17/29 coincide with the points $\beta$1,
$\beta$2 and $\beta$2$'$, $\beta$3. The families a14 and a14$'$ pass
through the point 29b of the spiral Eb and 29a of Ea, respectively. 
The lines Ea, Eb belong to the spiral structure around 1a as seen in
Fig.~\ref{figM1}.
(see also Fig.~3 in Barbanis 1993).

\begin{figure}
  \resizebox{\hsize}{!}{\includegraphics*{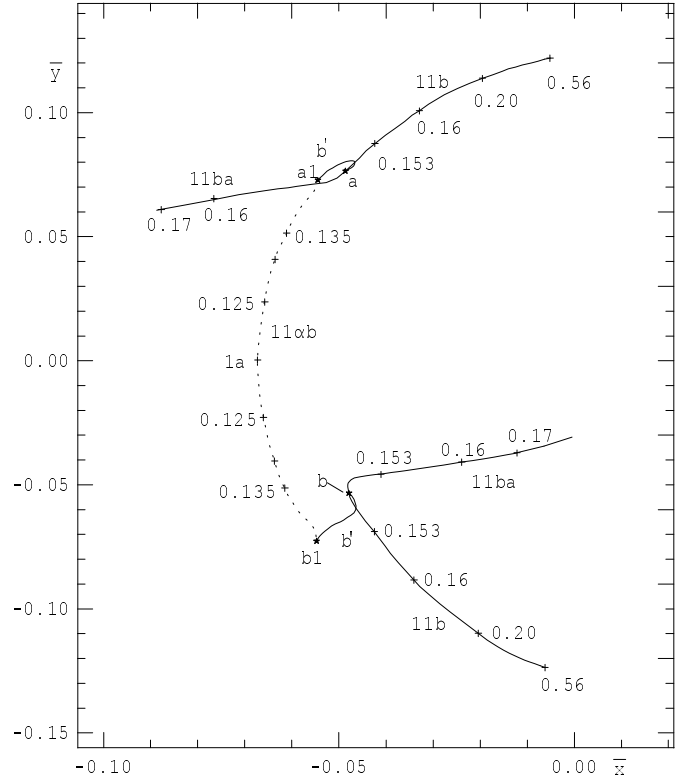}}
  \caption[]{The two branches of the family 11b connect to the branches of 
the family 11ba, at the points a, b. At these points emanate the branches 
of the family b$'$, which, by
changing $\eta$ from 0.20 to 0, reach the points a1, b1. At these points
terminate the branches of the family 11ab, which bifurcate from 1a when
$\epsilon=0.1227395$, $\eta=0$}
  \label{figA3}
\end{figure}

The families a18 and a18$'$ are connected indirectly with the families
$\beta$4 and $\beta$4$'$ respectively through two other families. A similar
example is shown in Fig.~\ref{figA3}. 

The family a19 bifurcates at $\epsilon_\mathrm{bif}=0.085467$ and exists for
$\epsilon < \epsilon_\mathrm{bif}$ until $\epsilon_\mathrm{min}=0.048765$. 
At this value it
is connected to a family which becomes highly unstable at
$\epsilon = 0.305$.

Each of the remaining orbits 29 and 29A of A , 29$'$ and 29A$'$ of
A$'$, c6 of C, 
c6$'$ of C$'$ belongs to a family which terminates at some minimum $\epsilon$,
where a new family emanates, so that six pairs are formed.
E.g. the orbit c6 is the 14th member of the basic sequence C.
This family terminates at $\epsilon_\mathrm{min}=0.53507$.
From this termination point emanates the family ca, which at $\epsilon=0.560$
reaches the point ca on a FTL (see Fig.~\ref{figA1}).
  
Summarizing, the bifurcating families evolve, with respect to
$\epsilon$, in the following way:
\begin{enumerate}
\item[a)] The characteristics of many families move away from the
original parent family and, for $\epsilon  = 0.560$, they cross a
BL or a FTL.
\item[b)] Some families terminate at a
maximum $\epsilon_\mathrm{max}$ or a minimum $\epsilon_\mathrm{min}$.
In this case there are three possibilities:
\begin{itemize}
    \item[(I)] Such a family is connected to another family
bifurcating also from the same orbit 1a or 1b and  terminating at
the same $\epsilon$ (see e.g. 11b2$' \to$11b8$'$, Table B1).
     \item[(ii)] From the same parent orbit, 1a or 1b, bifurcate two
families, one of them terminating at $\epsilon_\mathrm{max}$ and the
other at $\epsilon_\mathrm{min}$. Another family starting at
$\epsilon_\mathrm{min}$ terminates at $\epsilon_\mathrm{max}$ (see
Fig.~\ref{figA3}). In few cases there are more than one intervening families.
     \item[(iii)] From the bifurcating family, which stops
at $\epsilon_\mathrm{max}$, another family emanates which, for
$\epsilon = 0.560$, 
crosses a BL or a FTL (see e.g. 11b3 reaches 11f of F, F$'$,
 Table B1).
\end{itemize}
\item[c)] In bifurcating families with small rotation numbers the
computations stop sooner than $\epsilon = 0.560$, because the
orbits of these families become extremely unstable.
\end{enumerate}

\begin{figure}
  \resizebox{\hsize}{!}{\includegraphics*{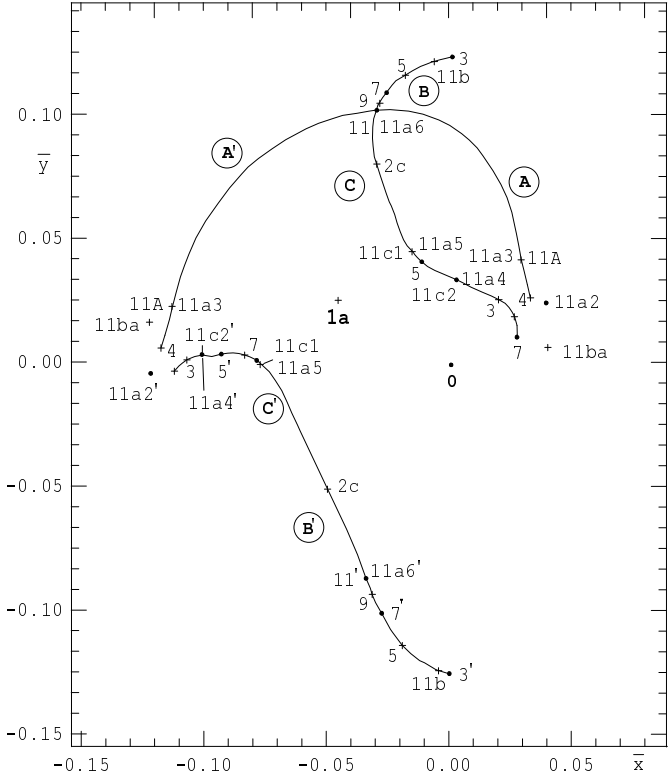}}
  \caption[]{The p.c.'s of the periodic orbits with m=11 are illustrated
with the corresponding bifurcating families from 1a. All the orbits on
  the BL's are connected directly with 1a, except of 11b, forming a
  pair with 11ba which is connected indirectly with 1a (see Fig~\ref{figA3})}
  \label{figA4}
\end{figure}

\begin{figure}
  \resizebox{\hsize}{!}{\includegraphics*{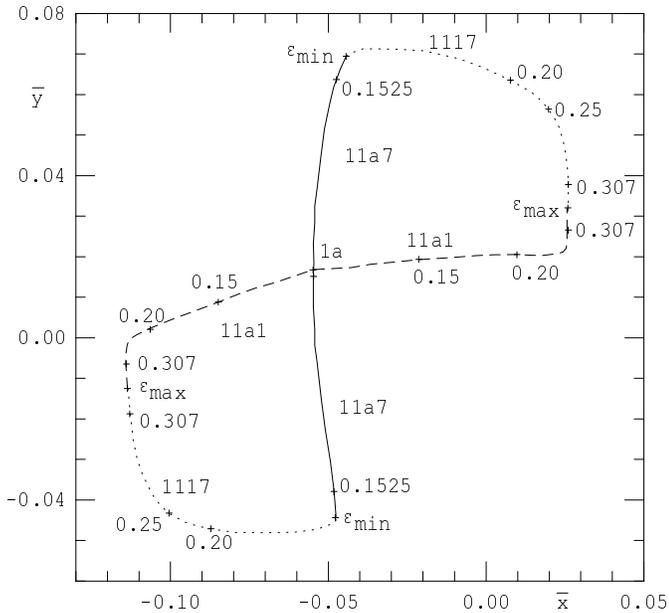}}
  \caption[]{The two branches of the bifurcating families 11a1 and 11a7
from 1a stop at $\epsilon_\mathrm{max} = 0.3099459$ and 
$\epsilon_\mathrm{min} = 0.1522655$ respectively. The family 1117 connects
these families.}
  \label{figA5}
\end{figure}

Some pairs of families that are connected at some $\epsilon_\mathrm{min}$
have no direct connection with 1a. The question is: Is there
some other family bifurcating from 1a, for some pair of values
of $\epsilon$ and $\eta$, which passes through the point from which
emanates the pair
when $\epsilon = 0.560$ and $\eta = 0.20$ ?

Fig.~\ref{figA2} represents the pair of 29A and 29Aa; their common point, S, 
corresponds
to $\epsilon_\mathrm{_1 min}=0.5504835$, $\eta=0.20$. Keeping
$\epsilon = \epsilon_\mathrm{_1 min} = const$ and varying $\eta$ we
find two other 
families, 29A1 and 29Aa1, emanating from the common point S. The family
29A1 stops at the  point, S1, corresponding to minimum $\eta_1 = 0.181232$, 
while
the family 29Aa1 stops at S2 when $\eta_2 = 0.184350$. All efforts to find
a continuation of 29Aa1 were unsuccessful. On the contrary, from the point
S1 emanate the families 29A2 and 29A3 for $\eta = \eta_1 = const$
and $\epsilon$ varying from the value $\epsilon_\mathrm{_1 min}$ to 
$\epsilon_\mathrm{_2 min} =
0.52840725$, where the two families terminate at the common point S3.
No connection of S3 to 1a has been found.

In the case of multiplicity 11 we find that all orbits
connect, directly or indirectly, to 1a, except for orbit 11b
with two p.c.'s (Fig.~\ref{figA4}). This orbit
is a Farey tree orbit on the lines B and B$'$ between the basic members 
3($\cdot$) and 5(+). 
The two branches of family 11b stop at
$\epsilon_\mathrm{_1 min} = 0.15204035$, $\eta = 0.20$ at the two
p.c.'s a, b (Fig.~\ref{figA3}).
From these points emanates the family 11ba which, for $\epsilon=0.560$, passes
through points 11ba (see Fig.~\ref{figA4}).
The two families form a pair with no direct connection to 1a. However,
there is an indirect connection to 1a. In fact, starting from the
common points a and b with constant $\epsilon = \epsilon_\mathrm{_1 min}$ 
and diminishing
$\eta$ from $\eta=0.20$ to $\eta=0$, we reach points a1 and
b1. Continuing this family, keeping $\eta=0$ and reducing now
$\epsilon$ until the value $\epsilon_\mathrm{bif} = 0.1227395$, we find
that the new family 11ab bifurcates from 1a. Therefore, families
11b and 11ba do not form a pair of irregular families, because this
pair is connected indirectly to 1a.

Since our results are numerical, it should be useful to confirm them through
further work for the following two reasons:
\begin{itemize}
\item[(I)] The search for bifurcations from 1a and 1b with multiplicity 29
has been confined within the interval $0 < \epsilon < 0.560$ and
$\eta=0.20$ only. 
One may argue that, although highly improbable,
bifurcations for $\epsilon > 0.560$ may play the role of connecting
bridges of the pairs, considered here as irregular families, to orbits
1a or 1b.  
\item[(ii)] Following the above argument as well as for reasons of
completeness, 
it would be desirable to find also the bifurcations of 1a
and 1b for $\epsilon = 0.560$ and $\eta \geq 0$.
\end{itemize}

     Table A2 shows the bifurcations of the orbits with
multiplicity m=11 from 1a when $0 \leq \epsilon \leq 0.560$, $\eta =
0.20$.

On the lines of Fig.~\ref{figA4} we represent the known orbits with
m=11, when $\epsilon=0.560$, $\eta=0.20$. These orbits are: 

\begin{table}
\caption[]{Bifurcations from 1a with m=11}
\begin{flushleft}
\begin{tabular}{lllll}
\hline\noalign{\smallskip}
R & $a$ & $\epsilon_\mathrm{bif}$ & Fam.  & Comments \\
\noalign{\smallskip}
\hline\noalign{\smallskip}
1/11 & -1.68250707 & 0.138524 & 11a1 & 11a1$\rightarrow$11a7 
\bigcr
 &  &  &  11a2 &   point 11a2 \smallcr
2/11 & -0.83083003 & 0.265946 &  & \smallcr
& & &  11a2$'$  & point 11a2$'$
\bigcr
3/11 & 0.28462968 & 0.373682 & 11a3  & 11a3=11A of
A, A$'$
\bigcr
& & & 11a4 &  11a4=11c2 of C\smallcr
4/11 & 1.30972147 & 0.463427 &  &\smallcr
& & & 11a4$'$ & 11a4$'$=11c2$'$ of C$'$
\bigcr
5/11 & 1.91898595 & 0.527056 & 11a5  & 11a5=11c1 of
C, C$'$
\bigcr
& & & 11a6   & 11a6=11 of B \& A\smallcr
6/11 & 1.91898595& 0.409869 &  &\smallcr
& & & 11a6$'$  & 11a6$'$=11$'$ of B$'$
\bigcr
7/11 & 1.30972147 & 0.155080 & 11a7  & 11a7$\rightarrow$11a1\\
\noalign{\smallskip}
\hline
\end{tabular}
\end{flushleft}
\end{table}

\begin{itemize}
\item On A and A$'$ : 11 (also on B), 11A
\item On B and B$'$ : 11 (also on A) and 11$'$, 11b
\item On C and C$'$ : 11c1, 11c2 and 11c2$'$
\item On other lines (not showing) 11a2 and 11a2$'$, 11ba
\end{itemize}

The bifurcations of Table A2 are related to the above orbits as
follows.

Family 11a1 is connected to family 11a7 through another
family 1117, as shown in Fig.~\ref{figA5}. Family 11a1 terminates at
$\epsilon_\mathrm{max} = 0.3099459$. Family  11a7 exists for 
$\epsilon \leq \epsilon_\mathrm{bif}$ and terminates at 
$\epsilon_\mathrm{min} =
0.1522655$. Family 1117 starts at $\epsilon_\mathrm{min}$ and
terminates at $\epsilon_\mathrm{max}$.

For $\epsilon = 0.560$, $\eta=0.20$ the following families pass through
the points illustrated in Fig.~\ref{figA4}
\begin{itemize}
\item Family 11a2 passes though point 11a2, lying on
a FTL (not shown). Similarly 11a2$'$  passes through
 11a2$'$
\item Family 11a3 passes through point 11A on A and A$'$. 
\item Family 11a4 reaches point 11c2 on C, while 11a4$'$ reaches
point 11c2$'$ on C$'$.
\item Family 11a5 reaches points 11c1 on  C and C$'$
\item Family 11a6 passes through the cross point 11 of
B and A, while 11a6$'$ reaches point 11$'$ on B$'$.
\end{itemize}

Families which start at points 11b on B and B$'$ and 11ba
form a pair when $\epsilon=0.15204035$, $\eta=0.20$. This pair has an
indirect connection to 1a, as illustrated in Fig.~\ref{figA3}.

\section{Search for irregular families; bifurcations from the orbit 1b}

Table B1 gives the bifurcations with multiplicity m=11 from 1b.
Families with $R=4/11$ to $7/11, 9/11$ and $10/11$ are not
included, because the corresponding values of $a$ are outside the
interval of the present study of $0 < \epsilon \leq 0.560$
and $\eta = 0.20$. 

\begin{figure}
  \resizebox{\hsize}{!}{\includegraphics*{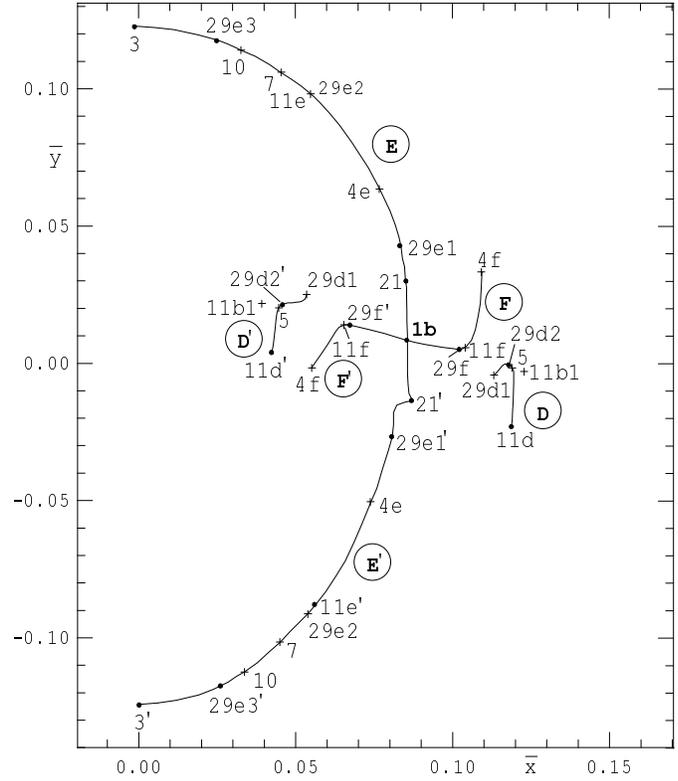}}
  \caption[]{The basic lines E and E$'$, F and F$'$ cross
each other at the periodic orbit 1b, while D and D$'$ lie close to 1b. On these
lines we show some basic orbits with small multiplicities and the
p.c.'s of the periodic orbits with m=11 and m=29.}
  \label{figB1}
\end{figure}

\begin{figure}
  \resizebox{\hsize}{!}{\includegraphics*{7919.fB2}}
  \caption[]{Location of the characteristics of the bifurcating families 
of Table B1 parametrized by $\epsilon$ ($\eta = 0.20$) until
 $\epsilon = 0.560$. Only the last points of family 11b3 are given.
The pair of the families 11e$'$ and 11d$'$ which
start through the corresponding orbits on E$'$ and D$'$ s also shown.
Family 11b8 is connected to 11d through the intervening family 118d.}
  \label{figB2}
\end{figure}

\begin{figure}
  \resizebox{\hsize}{!}{\includegraphics*{7919.fB3}}
  \caption[]{The bifurcating family 29b2$'$ is connected to 29b20$'$.
Families 29b2 and 29e3, corresponding to point 29e3 on E, are connected
indirectly through family 292e. Family 29b20 becomes highly unstable
at $\epsilon = 0.360$. Family 29e3$'$ forms a pair with 29x.  }
  \label{figB3}
\end{figure}

\begin{figure}
  \resizebox{\hsize}{!}{\includegraphics*{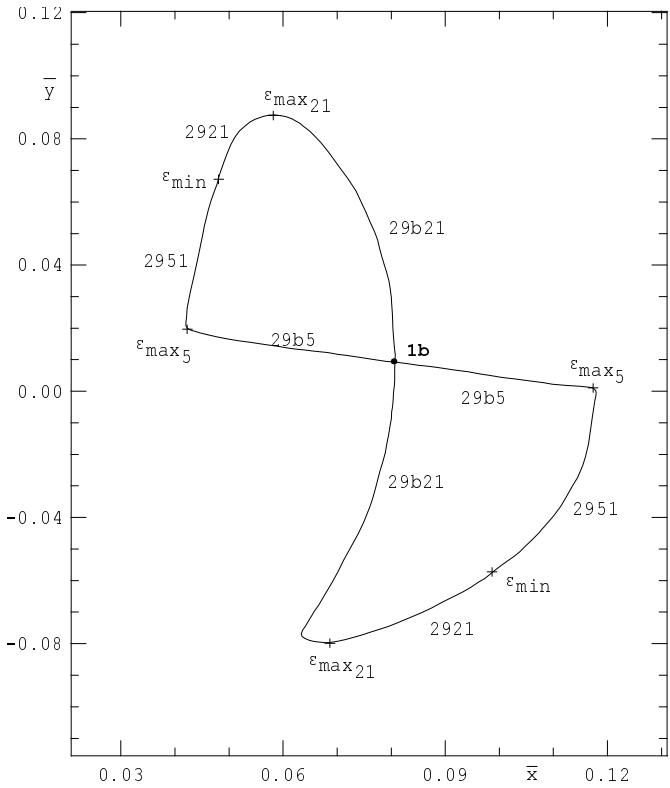}}
  \caption[]{The two branches of the bifurcating families 29b5 and 
  29b21 are connected through the intervening families 2951 and 2921.} 
  \label{figB4}
\end{figure}

\begin{figure}
  \resizebox{\hsize}{!}{\includegraphics*{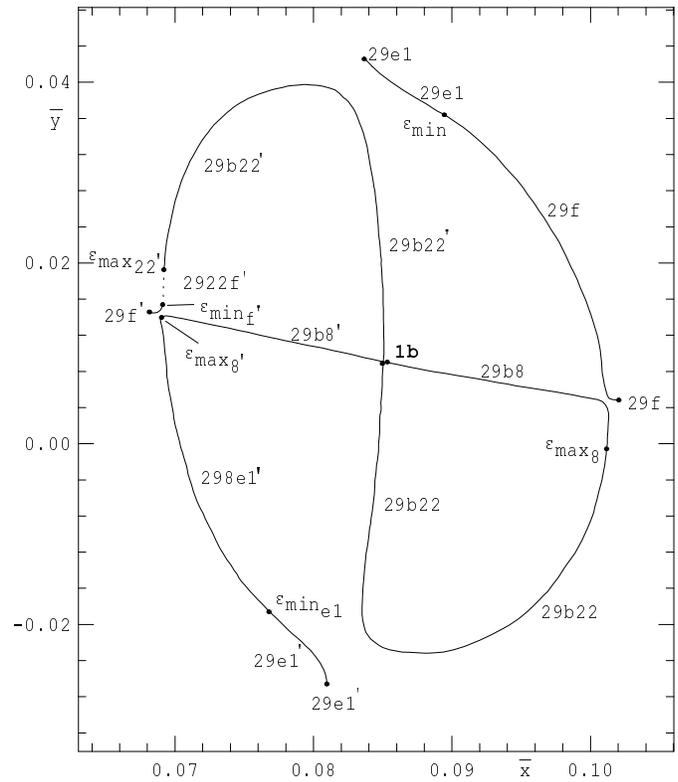}}
  \caption[]{The direct connection of 29b8 to 29b22 as well as the indirect
connections of 29b22$'$ to 29f$'$ through family 2922f$'$ and of 29b8$'$
to 29e1$'$ through family 298e1$'$ are illustrated. Families 29e1
and 29f, starting from points 29e1 on E and 29f on F respectively, form
a pair which has no obvious connection to 1b.}
  \label{figB5}
\end{figure}

\begin{figure}
  \resizebox{\hsize}{!}{\includegraphics*{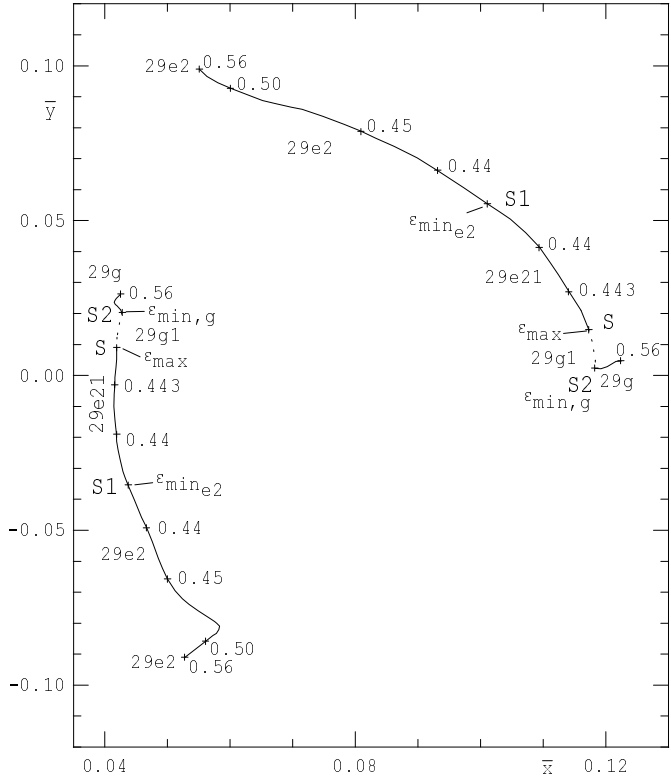}}
  \caption[]{Two pairs of families are connected at the common point
  S. The first pair, consisting of families 29e2 and 29e21, has as
  common point S1. The second pair of the families 29g and 29g1
 is connected at S2.}
  \label{figB6}
\end{figure}

\begin{table}
\caption[]{Bifurcations from 1b with m=11}
\begin{flushleft}
\begin{tabular}{lllll}
\hline\noalign{\smallskip}
R & $a$ & $\epsilon_\mathrm{bif}$ & Fam.  & Comments \\
\noalign{\smallskip}
\hline\noalign{\smallskip}
1/11 & -1.68250707 & 0.168452 & 11b1 & 11b1=11b1
\bigcr
& & & 11b2  & 11b2 = 11e of E\smallcr
2/11 & -0.83083003 & 0.344563 &  \smallcr
& & & 11b2$'$  & 11b2$'$ $\to$ 11b8$'$
\bigcr
3/11 & 0.28462968 & 0.525391 & 11b3 &  11b3=11f of
F, F$'$
\bigcr
& & & 11b8  & 11b8$\rightarrow$11d of D\smallcr
8/11 & 0.28462968 & 0.290248 &  \smallcr
& & & 11b8$'$  & 11b8$'$ $\to$ 11b2$'$\\
\noalign{\smallskip}
\hline
\end{tabular}
\end{flushleft}
\end{table}

\begin{table}
\caption[]{Bifurcations from 1b with m=29}
\begin{flushleft}
\begin{tabular}{lllll}
\hline\noalign{\smallskip}
R & $a$ & $\epsilon_\mathrm{bif}$ & Fam. & Comments \\
\noalign{\smallskip}
\hline\noalign{\smallskip}
1/29 & -1.95324111 & 0.063890 & 29b1 & highly unst.
\bigcr
&&&29b2 & 29b2$\rightarrow$29e3 \smallcr
2/29 & -1.81515084 & 0.127362 &  & \smallcr
&&&29b2$'$ & 29b2$'\to$29b20$'$
\bigcr
3/29 & -1.59218613 & 0.192211 &29b3  & highly unst.
\bigcr
&&&29b4 & \smallcr
4/29 & -1.29477257 & 0.258516 &  & $\gg$\smallcr
&&&29b4$'$ &  
\bigcr
5/29 & -0.93681688 & 0.326001 & 29b5  & 29b5$\rightarrow$29b21
\bigcr
&&&29b6 & 29b6 = 29d2  \smallcr
6/29 & -0.53505668 & 0.394292 & \smallcr
&&&29b6$'$ & 29b6$'$ =  29d2$'$
\bigcr
7/29 & -0.10827782  & 0.462973 & 29b7  & 29b7 = 29d1
\bigcr
&&&29b8 & 29b8$\rightarrow$29b22 \smallcr
8/29 & 0.32356399 & 0.531620 & \smallcr
&&&29b8$'$ & 29b8$'\rightarrow$29e1$'$
\bigcr
9/29 & 0.74027631 & $\epsilon_\mathrm{bif}$$>$$0.56$ 
\bigcr 
&&& 29b20 & 29b20 stops \smallcr
20/29 & 0.74027631 & 0.078653 & \smallcr
&&& 29b20$'$ & 29b20$'\to$29b2$'$
\bigcr
21/29  & 0.32356399 & 0.270218 & 29b21  & 29b21$\rightarrow$29b5 
\bigcr
&&&29b22 & 29b22$\rightarrow$29b28\smallcr
22/29 & -0.10827782 & 0.516148 & \smallcr
&&&29b22$'$ & 29b22$'$ $\rightarrow$29f$'$\\
\noalign{\smallskip}
\hline
\end{tabular}
\end{flushleft}
\end{table}

     Fig.~\ref{figB1} depicts the BL's E and E$'$, F and F$'$, D
and D$'$ and the places of the orbits with m=11 and m=29 when
$\epsilon=0.560, \eta=0.20$.

The known orbits on these lines are:
\begin{itemize}
\item [{a)}] m=11
  \begin {itemize}
\item On E and E$'$ : 11e , 11e$'$
\item On F on F$'$ : 11f
\item On D and D$'$ : 11d , 11d$'$d 
\end{itemize}
\item[{b)}] m=29
\begin{itemize}
\item On E and E$'$ : 29e1, 29e1$'$, 29e2, 29e3, 29e3$'$
\item On F and F$'$ : 29f, 29f$'$
\item On D and D$'$ : 29d1, 29d2, 29d2$'$
\end{itemize}
\end{itemize}

     Fig.~\ref{figB2} depicts the characteristics of the families in
Table B1, except for family 11b3, because its characteristic is
very close to that of family 11b1. Family 11b1 reaches
points 11b1 when $\epsilon = 0.560$, while family 11b3 
reaches the two points 11f of F, F$'$.

     Family 11b2 passes through point 11e of E.
Family 11b2$'$ reaches $\epsilon_\mathrm{max_{b8'}} = 0.46578405$,
where it is
connected to family 11b8$'$.  Family 11b8 reaches 
$\epsilon_\mathrm{max_{b8}} = 0.4622105$, where family 118d emanates.
This family connects 11b8 to 11d, which starts at $\epsilon_
\mathrm{min_d} =
0.443468$ and, when $\epsilon = 0.560$, it reaches point 11d
lying on D.

     Family 11d$'$ terminates at $\epsilon_\mathrm{min} = 0.443836$,
where it joins family 11e$'$, which for $\epsilon = 0.560$ passes through
point 11e$'$ of E$'$ (Fig.~\ref{figB2}). Families 11e$'$
and 11d$'$ form a pair, not having any obvious connection to the
basic orbit 1b.

     Table B2 shows the bifurcating families with multiplicities
m=29 from 1b.  As we mentioned before, the
computation of orbits
with small $R$, i.e. $R$ = 1/29, 3/29 and 4/29, stops before
$\epsilon = 0.560$ because these orbits become highly unstable.

     Fig.~\ref{figB3} shows the indirect connection of family 29b2
to the family 29e3, starting from orbit 29e3 on E.
Family 29b2 terminates at  $\epsilon_\mathrm{max_2}=0.3339236$, while
family 29e3 at $\epsilon_\mathrm{min}=0.2491375$. The family connecting
29b2 and 29e3 is 292e.

     Family 29b2$'$ is connected to 29b20$'$ at
$\epsilon_\mathrm{max_{2'}}=0.3334738$. On the other hand family 29b20
stops at $\epsilon=0.360$. Family 29e3$'$, starting from point 
29e3$'$ on E$'$, reaches $\epsilon_\mathrm{min}=0.2491375$, where
family 29x emanates. The computation of 29x stops at $\epsilon=0.408$,
$\eta=0.20$.

     Fig.~\ref{figB4} shows the connection of the branches of
families 29b5 and 
29b21 through two other families, i.e. 2951 and 2921.
Families 29b5 and 29b21 reach $\epsilon_\mathrm{max_5}=0.44357242$ and
$\epsilon_\mathrm{max_{21}}=0.4545552$ respectively. Families
2951 and 2921, which are
connected at $\epsilon_\mathrm{min}=0.4394852$, start,
respectively, in these points.

     For $\epsilon=0.560$ family 29b6 reaches point 29d2
of the FTL sequence 5(+), ....29d2($\cdot$), 24(+), 19($\cdot$),
14(+), 9($\cdot$)
on D, while 29b6$'$ is going through point 29d2$'$ on D$'$ 
(Fig.~\ref{figB1}).
Similarly for $\epsilon=0.560$ family 29b7 reaches points
29d1 of the basic sequence D and D$'$, namely the sequence with first
member the orbit 5(+) and increment 4.

Family 29b8 is
connected to 29b22 at $\epsilon_\mathrm{max_8}=0.55711905$ (Fig.~\ref{figB5}).
Family
29b8$'$ is connected indirectly to 29e1$'$, which passes through
point 29e1$'$ of E$'$ when $\epsilon=0.560$. The intervening family
298e1$'$ starts at $\epsilon_\mathrm{min_{e1}}=0.5520015$ where 29e1$'$ 
stops and
meets 29b8$'$ at its $\epsilon_\mathrm{max_{8'}}=0.556970$. Family 29b22$'$
continues until $\epsilon_\mathrm{max_{22'}}=0.55712395$, where family
2922f$'$ emanates. This  family  connects 29b22$'$
to family 29f$'$. Family 29f$'$ starts from orbit 29f$'$
on F$'$ when $\epsilon=0.560$ and stops at
$\epsilon_\mathrm{min_{f'}}=0.557062675$.

Families 29e1 and 29f, going through points 29e1 on E and 29f on F
respectively, form a pair connecting at $\epsilon_\mathrm{min}=0.5520015$.

The two branches of family 29e2, which start from points 29e2 on E 
and E$'$ when $\epsilon=0.560$, parametrized by $\epsilon$ ($\eta=0.20$)
stop at point S1 when $\epsilon_\mathrm{min_{e2}}=0.43790$
(Fig.~\ref{figB6}). A new family 29e21 emanates here  
and terminates at point S when
$\epsilon_\mathrm{max}=0.4437953$. This pair is connected 
to another pair which is formed by two families having two branches also.
The first family of this pair starts at $\epsilon=0.560$ and $\eta=0.20$ 
from points 29g, lying on FTL's and terminates at 
point S2, when $\epsilon_\mathrm{min,g}=
0.4435759$. At this point emanates the second family 29g1 of this pair.
This family terminates at S together with 29e21.
Such a double pair, having no connection to an orbit of the unperturbed
system, is noticed for the first time.

\end{document}